\documentclass{elsart}
\usepackage{amsmath}

\begin{document}
\runauthor{R.E. Cohen}
\begin{frontmatter}
\title{Theory of ferroelectrics: A vision for the next decade and beyond}
\author[]{Ronald E. Cohen}
\address{Carnegie Institution of Washington\\
5251 Broad Branch Rd., N.W.\\
Washington, D.C. 20015}

\begin{abstract}
In the last ten years significant advances have been made in our
understanding and ability to compute and predict properties of
ferroelectrics and piezoelectrics using fundamental physics. Phase
diagrams, electromechanical and elastic properties, and effects of
defects and surfaces are now amenable to computation. Most
importantly, new techniques have been developed, and new
understanding of the meaning of polarization in dielectrics has been
developed. Prospects for the future are discussed.
\end{abstract}
\end{frontmatter}

\section{Introduction}

About ten years ago a concerted effort began to apply modern
first-principles band structure methods to the problems of ferroelectricity,
with a vision that electromechanical properties could perhaps one day be
designed computationally \cite{1}. This effort has been at least
as successful as hoped, and a range of problems that we didn't even realize
existed at the time have been identified and solved. Understanding
electromechanical response has turned out to be a rich problem, both from a
fundamental condensed matter physics perspective, as well as from the
applied physics and even engineering perspectives. There is much expanded
interest in the field now that a new class of materials have been
discovered, single crystal piezoelectrics with ultra-high electromechanical
couplings such as PMN-PT (PbMg$_{\frac{1}{3}}$Nb$_{\frac{2}{3}}$O$_{3}$-PbTiO%
$_{3}$) which will revolutionize fields ranging from sonar and hydrophones
to medical ultrasonic imaging \cite{2}. The current state of the
field, as will be discussed below, is that we now can compute all of the parameters related to
electromechanical coupling, particularly the piezoelectric constants, for
pure, ordered, single crystal phases \cite{3,4}. This can be done with good 
accuracy at zero temperature, and
indications are that quite reliable results can also be obtained as
functions of temperature \cite{5}. The problems now being studied
are how to treat disordered solid solutions, and to understand the
underlying physics behind the observed material behavior--why is one
material better than another. In the future, using the computational
techniques developed, and the insights obtained from theory and experiment,
we will be able to predict material properties of as-yet unsynthesized
complex materials. The current state of the field and future prospects are
outlined here. This is not intended as a comprehensive review, but should be
a useful introduction to the field of first-principles studies of
ferroelectrics and piezoelectrics.

The reason to use first-principles methods is that (1) one is not reliant on
parameterized theories or (2) on fitting possible inaccurate experiments,
(3) one has access to the underlying potential surfaces, (4) can clearly see
the origin of observed behavior, and (5) they can be applied to hypothetical
or not yet synthesized materials, or (6) for temperature, pressure or
compositions for which data are not available. By ``first-principles,'' we
mean that experimental data are not used to constrain parameters; rather one
starts from the fundamental interactions among electrons and nuclei. Most of
the first-principles methods that have been applied to ferroelectrics are
based on the density functional theory (DFT) \cite{6} and
some are based on Hartree-Fock theory. The DFT states that the ground state
properties of a system are given by the charge density, and Kohn and Sham
\cite{7} showed how to compute the charge density and energy
self-consistently, using an effective exchange-correlation potential (V$%
_{xc} $) that accounts for the quantum mechanical interactions between
electrons. The local density approximation (LDA) takes the
exchange-correlation potential from the uniform electron gas at the density
for each point in the material. The Generalized Gradient Approximation (GGA)
includes the effects of local gradients in the density \cite{8}.
Given a form for the exchange correlation energy E$_{xc}[\rho]$, one
can find the self-consistent charge density and compute the energy for any
arrangement of nuclei (atoms). From the energies, zero temperature phase
diagrams, phonon frequencies, and elastic constants can be computed. In the
frozen phonon method, for example, one displaces atoms and computes the
change in energy as a function of displacement, from which the potential
surface for an instability, or phonon frequencies, can be derived. In
principle one can also compute finite temperature properties, for example
using Monte Carlo or molecular dynamics methods. In practice,
self-consistent methods are extremely computationally demanding, especially
for large systems, so instead one can find effective potentials or
Hamiltonians fit to the first-principles results to obtain finite
temperature properties.

Properties of ferroelectrics are extremely sensitive to volume
(pressure), which can cause problems since the present
first-principles methods are not perfect, and small errors in volume
(typically several percent, or more in some cases) can result in large
errors in computed ferroelectric properties. The GGA is generally an
improvement over the LDA, but still many computations for
ferroelectrics so far have been done at the experimental lattice.
The GGA gives reasonable $c/a$ at the experimental volume. For
PbTiO$_3$ at the experimental volume (427.27 bohr$^3$), for example,
we found $c/a$ equal to 1.12 (90\% high) for LDA (Hedin-Lundqvist),
1.09 for LDA (Wigner), and 1.073 for GGA (PBE), compared to the
experimental value of 1.063. However, the volume is still too large
in GGA and $c/a$ is thus overestimated. A new form for the exchange
and correlation called the Weighted Density Approximation (WDA)
shows much promise, and gives excellent results for the lattice
constants and energetics
\cite{9}, so there is hope that in the future all properties of
ferroelectrics could be computed truly
\textit{ab initio} without recourse to the experimental volume.

\section{Milestones}

\subsection{Total energy and electronic structure}

At first it was unknown whether first-principles methods within the LDA
would even give ferroelectric ground states where they should. If fact, in
the initial computations for BaTiO$_{3}$ (read ``initial'' as over one
year!) the author's LAPW computations did \emph{not }give a ferroelectric
ground state due to a minor bug that apparently did not effect other
systems! After this problem was fixed, the ferroelectric instability was
found at the experiment volume, though it was extremely sensitive to
pressure. The initial questions then after finding the instability were (1)
what is the underlying cause of the ferroelectric instability and (2) why do
different similar materials behave differently; particularly why does BaTiO$%
_{3}$ show a series of phase transitions from cubic, to tetragonal, to
orthorhombic, and to rhombohedral with decreasing temperature whereas PbTiO$%
_{3}$ has a single phase transition to tetragonal? The potential surface was
initially mapped out using the LAPW method, and the charge density and
electronic structure were analyzed 
\cite{10,11,12}.
PbTiO$_{3}$ showed a much deeper well when the tetragonal strain was
included, whereas in BaTiO$_{3}$ the rhombohedral phase had the lowest
energy. Thus the tetragonal strain is responsible for the tetragonal ground
state in PbTiO$_{3}$. Analysis of the charge densities and densities of
states showed that the ferroelectric instability is due to hybridization
between the O 2p states and the Ti 3d states, and if the 3d variational
freedom was removed from the problem, the ferroelectric instability
vanished \cite{10}. In BaTiO$_{3}$, the Ba is quite ionic and
spherical, whereas the Pb in PbTiO$_{3}$ is very not spherical in the
ferroelectric phase, and polarization of the Pb helps stabilize the large
strain and the tetragonal ground state in PbTiO$_{3}$.

Thus the following picture emerged. All Coulomb lattices are unstable with
respect to ferroelectric displacements; the short-range repulsive forces
tend to stabilize crystals with respect to off-center displacements. In
perovskite like BaTiO$_{3}$ and PbTiO$_{3}$, the O 2p states strongly
hybridize with the d$_{0}$ cation, in this case Ti$^{4+}$, reducing the
short-range repulsions thus allowing off-center displacements. Without
strain, or when strain effects are minor, the lowest energy off-center
displacements are along the (111) directions. Thus the ground state is
rhombohedral. As temperature is raised, the off-center displacements
disorder over two directions, giving an average orthorhombic symmetry, then
there is a disordering over four directions giving tetragonal symmetry, and
final a disordered cubic phase at high temperatures. The same features are
found in BaTiO$_{3}$ and KNbO$_{3}$ \cite{13,14}. One
should not overemphasize this order-disorder picture too much, however. In
the ideal order disorder picture, the atoms are in off-center sites
at all temperatures, and there should be a large configurational
entropy change at the ordering phase transitions, which is not
observed. Furthermore, in the ideal disorder picture, the local
potential for displacing a single B-site cation such as Ti or Nb is
unstable, but calculations show that the diagonal, self-force
constant, is positive, and so there would be a restoring force for
displacing a single cation from its ideal position in the perfect
cubic perovskite, and only when groups of ions are moved (for
example in a $\Gamma$-point displacement) is there a multiple well
potential surface \cite{15}.

It is clear that the polarizability of Pb plays a special role in
ferroelectrics. PbTiO$_{3}$ has a
very large strain (6\%) where tetragonal BaTiO$_{3}$ has only a 1\% strain.
The ground state of PbZrO$_{3}$ is antiferroelectric and it has a complex
phase diagram \cite{16}, whereas BaZrO$_{3}$ is cubic \cite{17}. PbZrO$_{3}$ has a 
complex structure at low temperatures, and it
was structural total energy relaxations within the LDA that clarified the
crystal structure \cite{16}. Thus first-principles theory has the
power now to find problems with experimental determinations of complex
structures.

Another example of theory finding problems with experimental analysis is
LiNbO$_{3}$. In the LiNbO$_{3}$ structure there are chains of oxygen
octahedra with Li ions shared between two octahedra. Based on neutron
scatter experiments, it was suggested that the Li atoms hopped between two
octahedra at high temperatures in the high temperature phase, and ordered to
occupied one set of octahedra at low temperatures in the polar
phase \cite{18,18b}. Analysis of experimental data suggested that the
underlying potential surface had a triple well structure for Li
displacements, and all models considered the Li motions to the important
physics driving the ferroelectricity. Inbar and Cohen performed LAPW frozen
phonon computations and found a quite different picture \cite{19}.
They found no triple well, and they found that the double well for Li
displacements was quite shallow, and the double well for oxygen displacements
much deeper. The deepest wells were found for coupled displacements of Li
and O. They found that the ferroelectric instabilities in LiNbO$_{3}$ and LiTaO%
$_{3}$ are quite similar to the instabilities in the oxide perovskites, and
that a primary driving force is the d$_{0}$ configuration of the Nb$^{5+}$
and Ta$^{5+}$ ions, and hybridization with the surrounding oxygens that
allows the d$_{0}$ cation to go off-center. This sets up local fields that
drive or order the Li off-centering.

Though static total energy computations are very powerful, and further
examples will be given below, they are limited in the ability to find phonon
or instability eigenvectors or behavior for arbitrary wavevector. Advances
that give the linear response for arbitrary wavevector are discussed next.

\subsection{First-principles lattice dynamics--linear response}

Using conventional methods, first-principles methods with the LDA or GGA
scale as N$^{3}$, where N is the size (say number of atoms) of the system.
This makes it prohibitive to study arbitrary wavevectors. For small
wavevectors a very large supercell would be required. In linear response
computations that problem is circumvented, and one can compute the response
of the system to small perturbations with arbitrary wavevector with
approximately the same cost as the primitive cell computation. Yu and
Krakauer developed a linear response LAPW code and applied it to phonon
dispersion in KNbO$_{3}$ \cite{13,20}. They found the
ferroelectric instability to be dispersive in certain directions, indicating
correlations in displacements in real space. In other words, if the unstable
band were flat, it wouldn't matter how the atoms were displaced in one cell
relative to the next. Analysis of the instability dispersion showed that
there should be chain-like correlations, consistent with streaking observed
in x-ray diffraction \cite{21}. Linear response methods also allow 
computation
of Born effective charges (discussed below) and the dielectric constant.

A number of different ferroelectric perovskites have now been studied using
linear response methods. The unstable modes of KNbO$_{3}$ and BaTiO$_{3}$
show very similar dispersion, consistent with the same sequence of phase
transitions for the two. The soft branch has primarily B-ion character. PbTiO%
$_{3}$ shows a much more flat soft branch, indicating that correlations
between cells are less important than in KNbO$_{3}$ or BaTiO$_{3}$. The soft
modes also have appreciable A-ion character (Pb) in PbTiO$_{3}$ compared
with KNbO$_{3}$ or BaTiO$_{3}$. PbZrO$_{3}$ is similar to PbTiO$_{3}$ in
that the soft branch remains unstable for any wavevector, but the zone
boundary R- and M-point instabilities are deeper than the $\Gamma $-point
instability, consistent with the fact that the ground state in PbZrO$_{3}$
is antiferroelectric, and PbTiO$_{3}$ is ferroelectric.

\subsection{Effective Hamiltonians--finite temperature properties}

Although self-consistent methods are very powerful, they generally cannot
provide anything but static, essentially zero temperature, properties, at
least for the complex systems we are discussing. Ferroelectric phase
transitions, however, occur with increasing temperature from one
ferroelectric state to another, or to the high temperature paraelectric
state. Dielectric and piezoelectric properties are strongly dependent on
temperature, especially around the phase transitions. Thus we want
techniques that can give finite temperature behavior. One approach is to use
a Gordon-Kim based model and Molecular Dynamics or Monte Carlo
\cite{22,23}, but these models have not been successful so far 
to
compute finite temperature properties for ferroelectrics. An effective
Hamiltonian approach has, however, been very successful \cite{24,25}. One develops 
a symmetrized expansion of the free
energy in terms of local mode coordinates, effective charges, and strains.
Only the important local modes are used, and thus the number is degrees of
freedom is reduced from the full atomic Hamiltonian. Local modes are
important if they are related to the soft mode, and the choice of local
modes is key to the success of the method. A rigorous formulation of local
modes as Wannier-like functions has been developed \cite{26}. Though
the effective Hamiltonian method contains all of the physics to give
qualitatively correct results, one might question whether it would be
quantitatively accurate since it neglects the higher frequencies modes, and
does not even try to estimate their average self-consistent effects on the
ferroelectric behavior. One such example is the importance of thermal
expansivity, which is not included in the effective Hamiltonian models so
far. Since ferroelectricity is very sensitive to volume, one would think
that neglect of thermal expansivity might lead to serious errors. This might
be partly responsible for the shifts in predicted transition temperatures
relative to experiment. So far, however, it appears, that the effective
Hamiltonian models are very accurate, and they are the only game in town for
finite temperature properties of ferroelectrics at the present time. Since
the underlying first-principles calculations are also not perfect, it is
surprising and encouraging that such high accuracy can be obtained.

\subsection{Physics of Polarization}

\subsubsection{Polarization in Periodic Boundary Conditions}

Contrary to intuition and many textbook discussions, polarization
cannot be determined directly from the self-consistent charge
density or changes in the charge density alone within periodic boundary
conditions. Rather one must consider changes in polarization as current
flow, and this can be determined from the phases of the wave functions.
King-Smith and Vanderbilt showed how this can be computed using a Berry's
phase approach \cite{27}. A more intuitive approach is the
realization that polarization changes can also be considered as the charge
transport on displacing Wannier functions \cite{28}. Practical
calculations generally use the Berry's phase approach, and the development
of this technique was a key advance in the ability to compute
electromechanical properties. A review of the theory of polarization
is given by Resta \cite{29}.

Born effective charges are defined as the change in polarization with
displacement of a nucleus, $Z_{i\alpha \beta }=dP_{\beta }/du_{i\alpha }$,
for the change in \ $\overrightarrow{P}$ moving ion $i$; note that the
effective charge is a tensor. The effective charges are not necessarily
equal to the nominal ionic charges, and can be quite different. In oxide
ferroelectrics they tend to be greatly enhanced for certain directions, with
values up to several times the nominal values. The
enhancement of the effective charges is due to strong covalent
hybridization \cite{30,31,32}.
Effective charges have been predicted for a number of ferroelectric systems
now. They are accessible to experimental determination, for example by
optical measurements of LO-TO splitting. Very large values were
experimentally obtained for BaTiO$_3$ many years ago \cite{33}, but there 
was no theoretical basis for understanding the results for many years.
It would be of great interest for experimentalists to measure these values
for a variety of ferroelectrics and compare with the theoretical predictions.

Piezoelectric constants can also be determined theoretically from changes in
polarization with respect to strain, as discussed below, using the Berry's
phase technique.

\subsubsection{Density polarization theory}

Another area of much discussion is in even more fundamental problems of
polarization under Kohn-Sham theory. It was argued that DFT
did not properly account for electronic polarization, and implied that
previous calculations were incorrect, and that the gap problem and low
dielectric constants in LDA were due to this failure \cite{34}. Mazin and 
Cohen
\cite{35} questioned this, and showed that there is no problem
with established methods such as the LDA. The situation has been clarified
\cite{36,37} and it appears that to obtain the
proper polarization in exact Kohn-Sham theory (which no one knows how to do
for real bulk systems of interest, anyway) one would have to consider the
polarization as well as the charge density. Realistic estimates of this
effect suggest it may be very small (3\%, \cite{38}). Regardless of the 
theoretical
importance of density polarization in the exact Kohn-Sham theory, it is
questionable if it is solely responsible for the gap problem and LDA
dielectric constants. The gap problem varies considerably from one material
to another, as do dielectric constants. For example, although the LDA\ high
frequency dielectric constant is 20\% too high in BaTiO$_{3}$, it is 
very close to experiment in PbTiO$_{3}$ \cite{39}. A study of a set of 11 
materials gives an r.m.s. error of 8\% from experiment for
$\epsilon_\infty$ using LDA \cite{40}. It is not at all clear than 
there is a uniform cause for these deviations. 

\subsection{Piezoelectricity}

The electromechanical response important in
transducer applications is given by the piezoelectric constants, or by the
figure of merit of the electromechanical coupling factors. Piezoelectric
constants can be written in different ways and one can transform from one
set to the other using thermodynamic transformations \cite{41}. The
set most easily computed are the piezoelectric strain coefficients $%
e_{ijk}=(\partial P_{i}/\partial \varepsilon _{jk})_{E=0}$, the change in
polarization with strain at zero applied field, usually reduced to the Voigt
form $\varepsilon _{j}$, $j=1...6$. Since the polarization can now be
computed using the Berry's phase approach, one can compute the polarization
versus strain and obtain the derivatives for the piezoelectric constants
numerically. It is important then to allow the atomic coordinates to
displace as functions of strain, because this is in fact the major
contribution to the piezoelectric response. The piezoelectric constant can
be written as

\begin{eqnarray}
e_{ijk} &=&\left( \frac{\partial P_{i}}{\partial \varepsilon _{jk}}\right)
=\left( \frac{\partial P_{i}}{\partial \varepsilon _{jk}}\right)
_{u}+\sum_{l\alpha }\left( \frac{\partial P_{i}}{\partial u_{l\alpha }}%
\right) _{\varepsilon }\left( \frac{\partial u_{l\alpha }}{\partial
\varepsilon _{jk}}\right) \\
&=&e_{ijk}^{h}+\sum_{l\alpha }\frac{e^{0}a_{i\alpha }}{\Omega }Z_{l,i\alpha
}^{\ast }\left( \frac{\partial u_{l\alpha }}{\partial \varepsilon _{jk}}%
\right) ,  \notag
\end{eqnarray}
where the first term, $e_{ijk}^{h}$, is the change of polarization with
homogeneous strain, that is no atomic relaxations, and the second is from
the displacements of the ions in response to strain, carrying effective
charges $Z$.

This was first carried through for PbTiO$_{3}$ \cite{3}.
Ferroelectric PbTiO$_{3}$ is tetragonal, and there are three independent
piezoelectric constants which can be found by three strains, tetragonal,
orthorhombic, and a monoclinic strain. Two methods can be used, the direct
method, that is finding the polarization versus strain, relaxing the
internal coordinates at each strain, and using the effective charges as
shown above. Both methods give the same final results, but the two step
procedure also illustrates the relative importance of the two terms.

First one must find the optimum minimum energy structure at the volume of
interest, in our case, the experimental zero pressure volume. Then, for
example for $e_{33}$(using Voigt notation), one varies the $c$-axis length,
and computes the change in polarization, giving $e_{33}^{h}$. One finds the
effective changes by displacing each ion separately, and computing the
change in polarization in each case, and then finds how the atoms relax as a
function of strain. Finally one can compute the polarization at the final
relaxed strained lattice. This gives two computations of $e_{33}$, which we
checked are identical, and give 3.23 C/m$^{2}$. The homogeneous contribution
is -0.88, and the internal relaxations give 4.11, so the total is dominated
by the atomic relaxations. Similarly, we found $e_{15}$=3.15 C/m$^{2}$ and $%
e_{31}$=-0.93 C/m$^{2}$. These are the proper moduli that do not include
homogeneous deformations of the original spontaneous polarization, which are
not observable. As pointed out later in this volume, the straightforward way
to extract the proper piezoelectric response is to compute differences in
Berry's phases, and convert to piezoelectric constants only in the last
step \cite{42}. The computed moduli were very close to some
experimental results, but there is a wide spread between different
experiments even for PbTiO$_{3}$, indicating the importance of obtaining
better values of experimental piezoelectric moduli under well-defined
electrical boundary conditions in order to test theory.

We also computed $e_{33}$ for two ordered structures of PZT50/50, near the
morphotropic phase boundary and obtained values significantly lower than
experimental measurements for PZT ceramics (single crystals were never made
successfully for PZT) extrapolated to low temperatures \cite{4}. 
This lead us to conclude either that (1) extrinsic effects are very
important even at low temperatures in PZT and/or (2) that the single crystal 
$e_{33}$ does not dominate the piezoelectric response in the poled direction
in PZT.

Using an effective Hamiltonian and Monte Carlo, Garcia and Vanderbilt
studied piezoelectric response of BaTiO$_{3}$ as a function of applied field
and temperature \cite{43}. They obtained good agreement with the
temperature dependence of the piezoelectric constants, and found a field
induced rhombohedral to tetragonal phase transition at very large fields.
Rabe and Cockayne used an effective Hamiltonian for \ PbTiO$_{3}$ and found d%
$_{33}$ as a function of temperature using Monte Carlo, also finding good
agreement with the experimental temperature dependence, which peaks strongly
at the ferroelectric phase transition \cite{5}.

\subsection{Solid solutions}

Most useful piezoelectrics are solid solutions rather than pure
ordered compounds.
This allows their properties to be tuned to meet engineering specifications.
But more than that, many complex solid solution ferroelectrics, for example,
the new single crystal piezoelectrics with giant electromechanical coupling,
like PMN-PT, have much enhanced properties relative to the pure compounds.
Even the endmember PMN is complex, with ions of different valence, Mg$^{2+}$
and Nb$^{5+}$ on the same crystallographic site. There has been some work to
study the ordering energetics in these systems. In PST-PT, for example, the
PIB (Potential Induced Breathing) model was used to study the relative
energetics of ordering, and to find an order-disorder phase diagram basic on
the Cluster Variation Method (CVM) \cite{44}. More recently,
self-consistent methods have been used along with Monte Carlo
\cite{45}. Even simple Madelung sums have been shown to illustrate some 
of the
important behavior in these systems \cite{46}. Only limited work so 
far
has studied how ordering (short-range and long-range) interacts with
ferroelectricity and piezoelectricity. Computations show that different
ordering schemes can drastically affect piezoelectric response in one
case \cite{47}, but how general this is not yet known.

\subsection{Defects, domain boundaries, and surfaces}

Real crystals are not perfect and infinite, and defects, domain boundaries,
and surfaces play important roles in ferroelectrics. Park and Chadi studied oxygen 
vacancies in PbTiO$_3$, which may be important in fatigue.  Domains have been 
studied using an effective Hamiltonian, and it was found that 180${{}^\circ}$ 
domain boundaries are very sharp, and even one unit cell away the
structure looks like bulk \cite{48} Large scale LAPW
computations were performed on periodic BaTiO$_{3}$ slabs. The
computations showed that just one unit cell away from the surface, the
charge density is indistinguishable from bulk BaTiO$_{3}$. There are two
types of surfaces with either TiO$_{2}$ or BaO terminations. On the TiO$_{2}$
surface the dangling Ti-bond relaxes back onto the surface, but atomic
relaxations are quite small. A surface state was found in the band
structure. LAPW computations for symmetrically terminated surfaces were
straightforward to interpret, and using results from slabs with two BaO and
two TiO$_{2}$ surfaces, an average unrelaxed surface energy of 920 erg/cm$^2$ was 
determined by comparing with the bulk BaTiO$_{3}$ energy \cite{49}. Using 
plane waves and pseudopotentials, Padilla and Vanderbilt found an average 
relaxed surface energy of 1260 erg/cm$^2$ \cite{50}. Hartree-Fock 
calculations using Gaussian basis sets gave 1690 erg/cm$^2$
(unrelaxed) \cite{51}. The differences in basis sets as well as Hartree-Fock 
versus LDA may be 
responsible for the differences. The effect of periodic slabs versus isolated 
slabs, as used in the Hartree-Fock study, amount to only about 3\% in the average 
surface energy for symmetric slabs. Clearly these are exploratory studies, and we 
cannot confidently state the exact surface energy for BaTiO$_3$ except all agree it 
is high, which explains the difficulty of  of cleavage in BaTiO$_3$ and the rough 
surfaces that form on fracture.

In the LAPW study, periodic asymmetric and polar slabs were also
studied \cite{49}. As discussed there, asymmetric or polar periodic slabs 
produce 
artificial
potential gradients across the slab which is like an external applied
field. This is an undesirable effect that makes it difficult to interpret
the results. Nevertheless, it was interesting that the polar slab has a
higher energy than the ideal slab, due to the unscreened depolarization
field. The field is large enough to cause the slab to be metallic with
densities of states overlapping from one side of the slab to the other. It
is much less metallic than would be the case in the rigid band picture,
however, with oxygen density of states piling up at the Fermi level. This is
due to the fact that O$^{2-}$ really wants to be closed shell, and not
metallic, even on surfaces.

In the pseudopotential study, symmetric slabs were studied, but a
polarization parallel to the slab was also considered \cite{50}. The 
isolated slab Hartree Fock study is the most straightforward to interpret, and they 
found results similar to bulk BaTiO$_{3}$. The use of isolated, rather than 
periodic, slabs, greatly facilities interpretation of results for asymmetric or 
polar slabs \cite{51}.

Padilla and Vanderbilt also studied SrTiO$_3$ surfaces and found a
very similar surface energy of 1360 erg/cm$^2$ compared with 1260 in
BaTiO$_3$. In SrTiO$_3$, they found without relaxation a value 14\%
higher, very close to the estimate of 12\% for the surface
relaxation energy in BaTiO$_3$ of Cohen \cite{49}.

\section{The Future: A Vision}

We see that in the last ten years significant advances have been made in
computing properties and understanding origins of ferroelectric behavior. We
now have the ability to predict ferroelectric instabilities,
electromechanical coupling, phonon dispersion and optical spectroscopy,
elasticity, pressure behavior, order-disorder, and some surface and defect
properties of structurally and chemically relatively simple pure compounds.
Now the challenge is to address problems in more complex solid solutions,
some that are structurally and/or chemically heterogeneous and some with
frequency dependent properties.

\subsection{Relaxor systems}

In relaxor systems the dielectric response has a broad peak as a function of
temperature, rather than a sharp peak in a normal ferroelectric, and a
frequency dependent response. The origin of this behavior is still
controversial, but is most likely due to heterovalent disorder. It will be a
challenge to compute this behavior from first-principles, but models can be
parametrized using first-principles results, as in the effective Hamiltonian
models discussed above, and then molecular dynamics or Monte Carlo
simulations could be used to simulate still rather small disordered systems.
To simulate much larger, mesoscopic, systems, it may be necessary to
parametrize models for interactions between nanoregions, leaving the atomic
domain all together. But this may be possible using only \textit{ab initio}
results.

\subsection{New piezoelectrics}

The discovery of new single crystal piezoelectrics with huge
electromechanical coupling \cite{2,52,53} 
are revolutionizing the field of piezoelectric transducers.
Theory did not play a role in discovering these materials, but it may play
an important role in showing how they work and helping design better
materials. The new materials are rhombohedral, with a morphotropic phase
boundary to a tetragonal phase nearby. The large response is d$_{33}$, which
is an elongation along the cubic (001) direction, and parallel to the applied
field. The rhombohedral phase though has polarizations locally about
(111). As the
the field is increased, the numbers of domains with polarizations that
project in the (0,0,-1) direction decrease, and there is a large
piezoelectric strain induced as the local polarizations are slowly rotated
up towards the (001) direction. At a critical field, a field induced phase
transition to true tetragonal symmetry occurs, and the piezoelectric
response changes and becomes more normal.  An idea of why the strain is so
large in these materials can be understood by considering PbTiO$_{3}$. PbTiO$%
_{3}$ is tetragonal, with a large 6\% strain at zero pressure and low
temperatures. There is no rhombohedral phase stable in PbTiO$_{3}$ at zero
or positive pressures, but if there was, one would see a small strain in
that phase, and one would see a giant strain as one went from the
rhombohedral to tetragonal phases. One can think of phases like PMN-PT\ as
chemically engineered ``rhombohedral PbTiO$_{3}$.'' It is clear that the
polarization of the Pb$^{2+}$ ion is key to this large strain
behavior \cite{12}, though a thorough understanding of exactly how
the Pb drives the strain has not yet been fully elucidated

\subsection{Materials by design}

The goal of this research program is to lead to the ability to
computationally design useful materials. We are clearly on that path, but it
will be a few years before one can practically hope to routinely design
materials by computer. On the other hand, it is also largely a matter of
having the right good idea. We understood the role of Pb in the strain of
PbTiO$_{3}$ some years ago, and if we smart enough to think of trying to
make ``rhombohedral PbTiO$_{3}$'' we might have found this large strain
effect before experiments. However, theoretical developments were just a few
years behind to realistically do this, because the first first-principles
computation of piezoelectric constants in pure tetragonal PbTiO$_{3}$ were
only done this last year, and it hasn't been that long since the Berry's
phase approach to computing polarization was discovered. As we make further
advances, however, in the next revolution perhaps theory will lead the way
rather than follow.

\begin{ack}

This work was supported by ONR grant N00014-92-J-1019. I would like to thank T. 
Egami, H. Fu, H. Krakauer, I. Mazin, K. Rabe, R. Resta, G. Saghi-Szabo, and D. 
Vanderbilt for helpful discussions. Computations were performed on the Cray J90-
16/4096 at the Geophysical Laboratory, supported by NSF EAR-9512627, The Keck 
Foundation, and the Carnegie Institution of Washington.
\end{ack}


\begin{thebibliography}{999}

\bibitem{1}L.L. Boyer, R.E. Cohen, H. Krakauer and W.A. Smith, First principles calculations for ferroelectrics -- A vision, {\em Ferroelec.\/} {\bf 111} (1990) 1-7.

\bibitem{2}T.R. Shrout, S. Park, C.A. Randall, J.P. Shepard, L.B. Hackenberger, D.J. Pickrell and W.S. Hackenberger,  Recent advances in piezoelectric materials, {\em Proc. SPIE - Int. Soc. Opt. Eng. (USA)\/}, SPIE-Int. Soc. Opt. Eng, (Adelaide, SA, Australia, 1997).

\bibitem{3}G. Saghi-Szabo, R.E. Cohen, and H. Krakauer, First-principles study of piezoelectricity in PbTiO$_3$, {\em Phys. Rev. Lett.\/} {\bf 80} (1998) 4321-4324.

\bibitem{4}G. Saghi-Szabo, R.E. Cohen, and H. Krakauer, First-principles study of piezoelectricity in tetragonal PbTiO$_3$ and PbZr$_\half$Ti$_\half$O$_3$, {\em Phys. Rev. B\/}, in press (1999).

\bibitem{5}K.M. Rabe and E. Cockayne,  Temperature dependent dielectric and piezoelectric response of ferroelectrics from first principles, in: R.E. Cohen, ed., {\em First-principles Calculations for Ferroelectrics: Fifth Williamsburg Workshop\/}, AIP, 1998) 61-70.

\bibitem{6}P. Hohenberg and W. Kohn, Inhomogeneous electron gas, {\em Phys. Rev.\/} {\bf 136} (1964) 864-871.

\bibitem{7}W. Kohn and L.J. Sham, Self-consistent equations including exchange and correlation effects, {\em Phys. Rev. A\/} {\bf 140} (1965) 1133-1140.

\bibitem{8}J.P. Perdew, K. Burke and M. Ernzerhof, Generalized gradient approximation made simple, {\em Phys. Rev. Lett.\/} {\bf 77} (1996) 3865-3868.

\bibitem{9}I.I. Mazin and D.J. Singh, Weighted density functionals for ferroelectric materials, in: R.E. Cohen, ed., {\em First-Principles Calculations for Ferroelectrics}, (American Institute of Physics,  1998) 251-264.

\bibitem{10}R.E. Cohen and H. Krakauer, Lattice dynamics and origin of ferroelectricity in BaTiO$_3$: Linearized augmented plane wave total energy calculations, {\em Phys. Rev. B\/} {\bf 42} (1990) 6416-6423.

\bibitem{11}R.E. Cohen and H. Krakauer, Electronic structure studies of the differences in ferroelectric behavior of BaTiO$_3$ and PbTiO$_3$, {\em Ferroelec.\/} {\bf 136} (1992) 65-84.

\bibitem{12}R.E. Cohen, Origin of ferroelectricity in oxide ferroelectrics and the difference in ferroelectric behavior of BaTiO$_3$ and PbTiO$_3$, {\em Nature\/} {\bf 358} (1992) 136-138.

\bibitem{13}R. Yu and H. Krakauer, First-Principles Determination of Chain-Structure Instability in KNbO$_3$, {\em Phys. Rev. Lett.\/} {\bf 74} (1995) 4067-4070.

\bibitem{14}D. Singh, Local density and generalized gradient approximation studies of KNbO$_3$ and BaTiO$_3$, {\em Ferroelec.\/} {\bf 164} (1995) 143-152.

\bibitem{15}P. Ghosez, E. Coackayne, U.V. Waghmare and K.M. Rabe, Lattice dynamics of BaTiO$_3$, PbTiO$_3$, and PbZrO$_3$: A comparative first-principles study, {\em Phys. Rev. B\/} {\bf in press} (1999) 

\bibitem{16}D.J. Singh, Structure and energetics of Antiferroelectric PbZrO$_3$, {\em Phys. Rev. B\/} {\bf 52} (1995) 12559-12563.

\bibitem{17}R.D. King-Smith and D. Vanderbilt, First-principles investigation of ferroelectricity in perovskite compounds, {\em Phys. Rev. B\/} {\bf 49} (1994) 5828-5844.

\bibitem{18}H.J. Bakker, S. Hunsche and H. Kurz, Quantum-mechanical description of the ferroelectric phase transition in LiTaO$_3$, {\em Phys. Rev. B\/} {\bf 48} (1993) 9331-9335.

\bibitem[]{18b}H.J. Bakker, S. Hunsche and H. Kurz, Coherent phonon polaritons as probes of anharmonic phonons in ferroelectrics, {\em Rev. Mod. Phys.\/} {\bf 70} (1998) 523-536.

\bibitem{19}I. Inbar and R.E. Cohen, Comparison of the electronic structures and energetics of LiTaO$_3$ and LiNbO$_3$, {\em Phys. Rev. B\/} {\bf 53} (1994) 1193-1204.

\bibitem{20}R. Yu and H. Krakauer, Linear-response calculations within the linearized augmented plane-wave method, {\em Phys. Rev. B\/} {\bf 49} (1994) 4467-4477.

\bibitem{21}R. Comes, M. Lambert and A. Guinier, The chain structure of BaTiO$_3$ and KNbO$_3$, {\em Solid State Comm.\/} {\bf 6} (1968) 715-719.

\bibitem{22}Z. Gong and R.E. Cohen, Molecular dynamics study of PbTiO$_3$ using non-empirical potentials, {\em Ferroelec.\/} {\bf 136} (1992) 113-124.

\bibitem{23}L.L. Boyer, H.T. Stokes and M.J. Mehl, Application of a Kohn-Sham-like formulation of the self-consistent atomic deformation model, {\em Ferroelec.\/} {\bf 194} (1997) 173-186.

\bibitem{24}W. Zhong, D. Vanderbilt and K.M. Rabe, Phase transitions in BaTiO$_3$ from first principles, {\em Phys. Rev. Lett.\/} {\bf 73} (1994) 1861-1864.

\bibitem{25}W. Zhong, D. Vanderbilt and K.M. Rabe, First-principles theory of ferroelectric phase transitions for perovskites: The case of BaTiO$_3$, {\em Phys. Rev. B\/} {\bf 52} (1996) 6301-6312.

\bibitem{26}K.M. Rabe and U.V. Waghmare, Localized basis for effective lattice Hamiltonians: Lattice Wannier functions, {\em Phys. Rev. B\/} {\bf 52} (1996) 13236-13246.

\bibitem{27}R.D. King-Smith and D. Vanderbilt, Theory of polarization of crystalline solids, {\em Phys. Rev. B\/} {\bf 47} (1993) 1651-1654.

\bibitem{28}D. Vanderbilt and R.D. King-Smith, Electric polarization as a bulk quantity and its relation to surface charge, {\em Phys. Rev. B\/} {\bf 48} (1993) 4442-4455.

\bibitem{29}R. Resta, Macroscopic polarization in crystalline
dielectrics: the geometric phase approach, {\em Rev. Mod. Phys.} {\bf 66} (1994) 899-915.

\bibitem{30}R. Resta, M. Posternak and A. Baldereschi, Towards a quantum theory of polarization in ferroelectrics, {\em Phys. Rev. Lett.\/} {\bf 70} (1993) 1010-1013.

\bibitem{31}M. Posternak, R. Resta and A. Baldereschi, Role of covalent bonding in the polarization of perovskite oxides: the case of KNbO$_3$, {\em Phys. Rev. B\/} {\bf 50} (1994) 8911-8914.

\bibitem{32}W. Zhong, R.D. Kingsmith and D. Vanderbilt, Giant LO-TO splittings in perovskite ferroelectrics, {\em Phys. Rev. Lett.\/} {\bf 72} (1994) 3618-3621.

\bibitem{33}J.D. Axe, Apparent ionic charges and vibrational eigenmodes of BaTiO$_3$ and other perovskites, {\em Phys. Rev.\/} {\bf 157} (1967) 429-435.

\bibitem{34}G. Gonze, P. Ghosez and R.W. Godby, Density-Polarization Functional Theory of the Response of a Periodic Insulating Solid to an Electric Field, {\em Phys. Rev. Lett.\/} {\bf 74} (1995) 4035-4038.

\bibitem{35}I.I. Mazin and R.E. Cohen, Notes on the static dielectric response function in the density functional theory, {\em Ferroelec.\/} {\bf 194} (1997) 263-270.

\bibitem{36}G. Ortiz, I. Souza and R.M. Martin, Exchange-correlation hole in polarized insulators: Implications for the microscopic functional theory of dielectrics, {\em Phys. Rev. Lett.\/} {\bf 80} (1998) 353-356.

\bibitem{37}D. Vanderbilt, Nonlocality of Kohn-Sham exchange-correlation fields in dielectrics, {\em Phys. Rev. Lett.\/} {\bf 79} (1997) 3996-3969.

\bibitem{38}X. Gonze, P. Ghosez and R.W. Godby, Density-functional theory of polar insulators, {\em Phys. Rev. Lett.\/} {\bf 78} (1997) 294-297.

\bibitem{39}P. Ghosez, J.P. Michenaud and X. Gonze, The physics of dynamical atomic charges: the case of ABO$_3$ compounds, {\em Phys. Rev. B\/} {\bf 58} (1998) 6224-6240.

\bibitem{40}F. Bernardini and V. Fiorentini, Electronic dielectric constants of insulators calculated by the polarization method, {\em Phys. Rev. B\/} {\bf 58} (1998) 15292-15295.

\bibitem{41}M.E. Lines and A.M. Glass, {\it Principles and Applications of Ferroelectrics and Related Materials\/} (Clarendon Press, Oxford, 1977).

\bibitem{42}D. Vanderbilt, {\em this volume\/}  (1999).

\bibitem{43}A. Garcia and D. Vanderbilt, Electromechanical behavior of BaTiO$_3$ from first principles, {\em Appl. Phys. Lett.\/} {\bf 72} (1998) 2981-2983.

\bibitem{44}B.P. Burton and R.E. Cohen, Non-empirical calculation of the Pb(Sc0.5Ta0.5)O$_3$-PbTiO$_3$ quasibinary phase diagram, {\em Phys. Rev. B\/} {\bf 52} (1995) 792-797.

\bibitem{45}B.P. Burton, {\em this volume\/} (1999).

\bibitem{46}L. Bellaiche and D. Vanderbilt, Electrostatic model of atomic ordering in complex perovskite alloys, {\em Phys. Rev. Lett.\/} {\bf 81} (1998) 1318-1321.

\bibitem{47}E. Cockayne and K.M. Rabe, Enhancement of piezoelectricity in a mixed ferroelectric, {\em Phys. Rev. B\/} {\bf 57} (1998) R13973-R13976.

\bibitem{48}J. Padilla, W. Zhong and D. Vanderbilt, First-principles investigation of 180 (degrees) domain walls in BaTiO$_3$, {\em Phys. Rev. B\/} {\bf 53} (1996) R5969-R5973.

\bibitem{49}R.E. Cohen, Surface effects in ferroelectrics: Periodic slab computations for BaTiO$_3$, {\em Ferroelec.\/} {\bf 194} (1997) 323-342.

\bibitem{50}J. Padilla and D. Vanderbilt, Ab initio study of BaTiO$_3$ surfaces, {\em Phys. Rev. B\/} {\bf 56} (1997) 1625-1631.

\bibitem{51}L. Fu, E. Yashenko, L. Resca and R. Resta, Hartree-Fock studies of the ferroelectric perovskites, in: R.E. Cohen, ed., {\em First-Principles Calculations for Ferroelectrics: Fifth Williamsburg Workshop}, (AIP,  1998) 107-117.

\bibitem{52}S.E. Park and T.R. Shrout, Relaxor based ferroelectric single crystals for electro-mechanical actuators, {\em Materials Research Innovations\/} {\bf 1} (1997) 20-25.

\bibitem{53}K. Uchino,  High electromechanical coupling
piezoelectrics: relaxor and normal ferroelectric solid solutions, {\em Solid State Ion. Diffus. React. (Netherlands)\/}, Elsevier, (Chiba, Japan, 1997).


\end{thebibliography}
\end{document}